# An experimental study of partial melting and fractional crystallization on the HED parent body

Helen O. ASHCROFT* and Bernard J. WOOD

Department of Earth Sciences, University of Oxford, South Parks Road, Oxford OX1 3AN, UK
*Corresponding author. E-mail: helen.ashcroft@earth.ox.ac.uk

**Abstract**–We have performed an experimental and modeling study of the partial melting behavior of the HED parent body and of the fractional crystallization of liquids derived from its mantle. We estimated the mantle composition by assuming chondritic ratios of refractory lithophile elements, adjusting the Mg# and core size to match the density and moment of inertia of Vesta, and the compositions of Mg-rich olivines found in diogenites. The liquidus of a mantle with Mg# (=100*[Mg/(Mg+Fe)]) 80 is ~1625 °C and, under equilibrium conditions, the melt crystallizes olivine alone until it is joined by orthopyroxene at 1350 °C. We synthesized the melt from our 1350 °C experiment and simulated its fractional crystallization path. Orthopyroxene crystallizes until it is replaced by pigeonite at 1200 °C. Liquids become eucritic and crystal assemblages resemble diogenites below 1250 °C. MELTS correctly predicts the olivine liquidus but overestimates the orthopyroxene liquidus by ~70 °C. Predicted melt compositions are in reasonable agreement with those generated experimentally. We used MELTS to determine that the range of mantle compositions that can produce eucritic liquids and diogenitic solids in a magma ocean model is Mg# 75–80 (with chondritic ratios of refractory elements). A mantle with Mg# ~ 70 can produce eucrites and diogenites through sequential partial melting.

## INTRODUCTION

The HED meteorites are volcanic and plutonic rocks believed to represent the crust and mantle of the asteroid Vesta. The eucrites sample the shallow crustal lava flows and associated intrusive rocks on Vesta and are split into subgroups depending on their texture and bulk composition. They contain roughly equal proportions of Ca-rich plagioclase ($An_{96-75}$) and pigeonite, with minor olivine, chromian spinel, silica minerals, ilmenite, FeNi metal, troilite, and phosphates (McSween et al. 2011). Diogenites are orthopyroxene-rich cumulates subdivided into orthopyroxenites and harzburgites (Beck et al. 2013) containing orthopyroxene ($Wo_{1-3}En_{71-77}Fs_{22-24}$), olivine, chromite, phosphates, metal, and occasional plagioclase (McSween et al. 2011). Howardites are polymict breccias containing both eucrite and diogenite fragments, together with other fragments including CM CI and CR chondrite material (Zolensky et al. 1996). Although the exact links between the eucrites and diogenites are incompletely understood, the homogeneous oxygen isotope signatures of the HEDs (Greenwood et al. 2014) and their similarities in Fe/Mn ratio (Papike et al. 2003) indicate that they are derived from a single planetary body that has experienced extensive melting. The recent compositional mapping of the surface of Vesta (e.g., Ammannito et al. 2013), showed a heterogeneous surface with distinct regions of each of the HED lithologies.

Petrogenetic models for the origin of eucrites and diogenites initially invoked conventional principles of equilibrium partial melting and fractional crystallization. Stolper (1977), for example, suggested that a eucrite source region comprised of olivine (~$Fo_{65}$), pyroxene (~$Wo_5En_{65}$), plagioclase (~$An_{94}$), Cr-rich spinel, and metal underwent various degrees of partial melting to produce liquids approximated by Stannern and Sioux County eucrites. Other eucritic compositions could then, he argued, be generated by low-pressure fractionation of liquids similar in composition to Sioux County. In this model, the magnesian pyroxenes and rare olivines of diogenites and howardites crystallized from melt produced after plagioclase was exhausted from the eucrite source regions. The alternative hypothesis (Ikeda and Takeda 1985) invokes global melting of the HED parent body followed by crystal accumulation in the magma ocean to generate diogenites, with noncumulate eucrites being formed by subsequent crystallization of the residual melts (Ruzicka et al. 1997; Takeda 1997). This magma ocean model was extended by Righter and Drake (1997) who proposed that large-scale melting of the HED parent body was accompanied by metallic core segregation and silicate crystallization in two stages. The first stage involved ~80% of equilibrium crystallization. The remaining ~20% melt segregated when the fraction of crystals became too high for the magma ocean to convect efficiently. The segregated melt then underwent fractional crystallization to produce differentiated eucrites as the melts and crystalline diogenites as the fractionating crystals. More recently, Mandler and Elkins-Tanton (2013) used the MELTS software (Ghiorso and Sack 1995) to test this model and argued for a lesser extent of equilibrium crystallization of the magma ocean (60–70%) before melt segregation.

Although useful starting points, neither the partial melting nor the magma ocean model accounts for the entire range of lithologies and compositions found in the meteorite record. The partial melting or serial magmatism hypothesis can account for the large variations in major and trace element abundances seen in eucrites and diogenites given more than one source composition and multiple melting events (e.g., Barrat et al. 2000). The residual liquids after diogenite crystallization are not, however, represented by meteorites (Mittlefehldt et al. 2012). The magma ocean model can crystallize general eucrite and diogenite compositions from one source composition but cannot account for the observed ranges in trace element abundances particularly in the diogenites (Beck and McSween 2010). Furthermore, thermochemical modeling of asteroidal melting (Wilson and Keil 2012; Neumann et al. 2014) suggests that Vesta may not have been able to sustain a global magma ocean. In contrast, either the magma ocean or partial melting model could be consistent with W and Mg-isotopic observations indicating that diogenites crystallized before eucrites (Schiller et al. 2011).

Although the thermal and magmatic histories of Vesta are uncertain, there is widespread evidence for the thermal metamorphism of the eucrites (Yamaguchi et al. 1996). Crustal anatexis produced partial melts which were incorporated into erupting eucrite magmas, a process whose trace element signature is seen in the Stannern Trend eucrites (Barrat et al. 2007) and which left behind depleted granulitic eucrites (Yamaguchi et al. 2009). Similar partial melting of early ultramafic cumulates may also have generated diogenites (Barrat and Yamaguchi 2014).

Any petrogenetic model of Vesta hinges, of course, on its bulk composition and on the composition of its silicate envelope. Bulk silicate Vesta appears to be like Earth, Mars, and the Moon in that the refractory lithophile elements are in approximately chondritic proportions (Morgan et al. 1978). For this reason, chondrite-like compositions are often used as putative starting bulk compositions for the HED parent body, and Righter and Drake (1997) and Toplis et al. (2013) have both explored the crystallization paths of a wide range of different chondritic starting compositions using the MELTS program (Ghiorso and Sack 1995).

Righter and Drake (1997) derived a range of starting compositions by using the abundances of moderately siderophile elements to model core formation and estimate a mantle composition. A 30:70 mixture of CV and L type chondrites was selected as an optimal composition for their equilibrium and fractional crystallization models; this satisfies the Fe/Mn ratios seen in HED meteorites and their oxygen isotopes. The molar Mg# (=100*[Mg/(Mg+Fe)], atomic) of this starting composition was 74 and equilibrium and fractional crystallization simulations were performed over the 1530–1150 °C temperature range. The major element starting composition is similar to the one we have derived (see below), but with a lower Mg# (Table 1).

Although the calibration database for the MELTS software includes studies of HED compositions (Stolper 1977; Longhi and Pan 1988; Bartels and Grove 1991; Grove and Bartels 1992) and of other relevant chondritic compositions (Jurewicz et al. 1991, 1993), modeling of partial melting and crystallization of the HED parent body requires extrapolation outside of the compositionally calibrated range. The issue with this is that the accuracy or amount of error cannot be estimated and a recent study by Balta and McSween (2013) highlighted several issues with the applicability of MELTS to similar Martian compositions. Therefore one important task in developing a petrogenetic model of the HED parent body is to test the accuracy of MELTS to model eucrite and diogenite formation from putative mantle compositions. That is one of the major aims of this study. A second aim is to test the validities of different published models of eucrite-diogenite relationships: the two-stage model of a global magma ocean (Righter and Drake 1997), and partial melting-fractional crystallization models. In order to do this, we estimated the composition of the Vestan mantle and

Table 1. Bulk silicate Vesta compositions (wt%).

|  | PM2 | RD |
|---|---|---|
| $SiO_2$ | 44.83 | 46.01 |
| $TiO_2$ | 0.58 | 0.15 |
| $Al_2O_3$ | 3.59 | 3.11 |
| $Cr_2O_3$ | 0.24 | 0.67 |
| FeO | 14.26 | 18.18 |
| MgO | 31.02 | 29.25 |
| CaO | 2.99 | 2.49 |
| $Na_2O$ |  | 0.11 |
| $K_2O$ |  | 0.01 |
| MnO | 0.41 |  |
| NiO | 0.52 |  |
| Mg# | 79.5 | 74.2 |

PM2 = the bulk silicate Vesta composition derived here. Minor NiO added to examine trace element partitioning (see text).
RD = Righter and Drake (1997) 0.3 CV—0.7 L composition with depleted volatiles.

performed a series of partial melting experiments. Analysis of the products was then followed by several sequences of crystallization experiments with the aim of being able to model fractional crystallization of parental eucritic melts. We begin with estimates of the composition of Vesta's mantle.

## ESTIMATES OF COMPOSITION OF THE HED PARENT BODY MANTLE

The recent Dawn mission produced a high-resolution shape model of Vesta and a description of its gravitational field yielding values of a bulk density of 3456 kg m$^{-3}$ and a gravitational moment of inertia of 0.03178 (Russell et al. 2012). The moment of inertia measurements indicate that Vesta has a core, although there is no unique solution regarding the size and density of the interior layers.

Applying the geophysical constraints together with the assumption of chondritic refractory element ratios provides the starting point for our model of Vesta. An additional constraint we applied was that the mantle be able to generate both eucrite and diogenite lithologies. Given this, an Mg# of 80 was selected as a starting point for the mantle composition, from the most Mg-rich olivines sampled in diogenites (Beck et al. 2012). The abundances of the other major element components in silicate Vesta were subsequently calculated from chondritic ratios. The starting composition is shown in Table 1 and is referred to as PM2 from here onward.

An interior model of the HED parent body was calculated in order to check the plausibility of the putative mantle composition. Vesta can be modeled as a triaxial ellipsoid with three interior layers (crust, mantle, core). Equations for the bulk density ($\rho_{bulk}$) and polar moment of inertia ($I_{zz}$) are then derived as follows:

$$I_{zz} = \frac{8\pi}{15}\left[(\rho_{crust}a_1^4 c_1) + ((\rho_{mantle} - \rho_{crust})a_2^4 c_2) + ((\rho_{core} - \rho_{mantle})a_3^4 c_3)\right] \quad (1)$$

$$\rho_{bulk} = (\rho_{crust}a_1^2 c_1) + ((\rho_{mantle} - \rho_{crust})a_2^2 c_2) + ((\rho_{core} - \rho_{mantle})a_3^2 c_3) \quad (2)$$

where $a_1$ and $c_1$ are the major and minor axes of the body, $a_2$ and $c_2$ are the mantle radii, $a_3$ and $c_3$ the core radii, and the $\rho$ denotes the density of each layer. We assume that the core is FeNi metal with a density between 7100 and 7800 kg m$^{-3}$, an assumption supported by the depletions of siderophile elements in the HED meteorites (Righter and Drake 1997). An average density of eucrites and diogenites taken from Britt et al. (2010) was used to estimate a crustal density of 2900 kg m$^{-3}$, while $I_{zz}$ was calculated for a range of crustal thicknesses (20–50 km) and core densities (7000–8000 kg m$^{-3}$). We obtained a polar moment of inertia between 0.4 and 0.42 (comparable with the 0.38–0.42 of the two layer model calculated by Rambaux [2013]). The calculated core masses are 15–20% by mass of Vesta, with mean core radii (assuming a spherical core) of ~110–120 km. This estimate is comparable to that of Russell et al. (2012), who estimate a core that is 18% by mass of Vesta with a radius of 107–113 km and Park et al. (2014)'s fit to Vesta's gravity model, which produced a spheroidal core (117 × 105 km) which is 17% by mass of Vesta. In addition, our result is comparable with the results of the geochemical modeling of core-mantle partitioning of the moderately siderophile elements by Righter and Drake (1997) which indicate a core of 15–25% of Vesta's mass. Although these solutions are not unique, and the calculations are simplified and do not account for the possible porosities of different layers, these comparisons indicate that our putative mantle composition is a reasonable starting composition.

## EXPERIMENTS

Starting compositions were prepared from mixtures of analytical grade oxide and carbonate powders. A small amount of NiO was added (Table 1) with the intention of measuring crystal-liquid partitioning of Ni. A significant fraction of this was lost to the Re wire, however, during the experiments. The oxides were fired overnight at 1100 °C to remove $CO_2$ and $H_2O$. The mixtures were ground under ethanol in an agate mortar and partially decarbonated by heating from 500 to 900 °C over several hours. They were then reground, pelletized, and reduced overnight at 900 °C in a vertical gas-mixing furnace at an $fO_2$ just above the IW buffer. After reduction, the pellets were reground under ethanol, before being pressed into pellets of around 500 mg, from which chips were broken off and used in the experiments.

Given the low pressures inside Vesta (~2 kbar in its center) we are justified in exploring petrogenetic relationships in outer silicate Vesta at atmospheric pressure. This approach also enables direct oxygen fugacity control under strictly anhydrous conditions. Samples were suspended on a loop of 0.25 mm diameter Re wire in a flowing $CO-CO_2$ gas mixture to control the oxygen fugacity. At the end of the experiment the sample was drop-quenched into water. The $fO_2$, temperature, run duration, and phases present (and their modal proportions) in each experimental charge are reported in Table 2 and the major element compositions of each phase are reported in Table S1 together with their uncertainties.

In order to determine the liquid line of descent and the compositions of mantle partial melts, crystallization experiments were performed on PM2 at 50 °C temperature intervals between 1300 and 1650 °C and an $f_{O2}$ 1.8 log units above the IW buffer. This $fO_2$ was selected to ensure a very high $Fe^{2+}/Fe^{3+}$ ratio in the melt, no metallic Fe precipitation, and no significant iron loss to the Re wire. The 50 °C intervals between experimental temperatures were selected in order to monitor the evolution of the liquid composition at a resolvable temperature interval, with the expectation that no major phases would be missed in the crystallization sequence.

In order to simulate fractional crystallization, or the second stage of possible magma ocean evolution, the glass composition observed at 1350 °C (after about 55% equilibrium crystallization and the appearance of orthopyroxene) was synthesized (HA4). A sequence of crystallization experiments was performed with this melt composition from 1350 to 1250 °C. The glass composition in the 1250 °C experiment was then synthesized (HA5) and crystallization experiments were performed between 1250 and 1150 °C. The melt produced at 1150 °C (HA7) was then separated, synthesized, and crystallized down to 1125 °C.

## ANALYSIS AND RESULTS

All phases present in the experimental products were analyzed in WDS-mode using a JEOL JXA-8600 electron microprobe in the Research Laboratory for Archeology and the History of Art at the University of Oxford. All phases were analyzed using a 15 kV accelerating voltage, with a 10 nA beam current and a focused beam. Counting times were typically 30–50 s on peak and 15–25 s on backgrounds. Natural and synthetic standards were used for calibration and during analytical sessions an almandine crystal was analyzed together with the experimental charges to check calibration and ensure internal consistency. All data were subsequently reduced using a PAP procedure (Pouchou and Pichoir 1991). Crystal analyses were accepted if their totals were between 98–101 wt% and glass analyses with totals over 97 wt% were generally accepted because the glasses typically contain up to 1 wt% trace elements.

### Attainment of Equilibrium

Several approaches were taken in order to determine whether equilibrium was reached and whether Fe was being lost to the Re loop. Firstly the phase compositions from experiment HAPM2_05 were synthesized separately and mechanically mixed together in a 70:15:15 ratio of melt:pyroxene:olivine to form the HA3 composition. These proportions were chosen in order to grow fewer and larger crystals for better analysis. A time series of experiments was performed using this composition for 24 (HA3_01) and 48 (HA3_03) hours, and two experiments were run on the same Re loop (HA3_01 and HA3_02). Experiments where the charge was heated up to the run temperature from 800 °C (e.g., Experiment HA3_01), as well as an experiment where the charge was heated from 800 to 1400 °C before being cooled to 1350 °C (HA3_06) were performed. From the major element analyses (Table S1) it can be seen that the phase proportions and Mg# of each experimental charge are all within uncertainty of each other, indicating that equilibrium was approached. Although no detectable Fe was lost from the charges, we found progressive Ni loss with increasing experiment duration.

### Phase Proportion Calculations

The calculation of modal proportions in each experiment from mass balance employed the nonlinear least squares approach of Albarede and Provost (1977). This method involves iteratively calculating a bulk composition from the individual phase compositions by varying the modal proportions of the phases, until the differences between the measured and calculated compositions have been minimized. The results are shown in Table 2. These calculations are sensitive to minor changes in phase composition that account for the minor differences between the calculated modal proportions for HAPM2_05 (45% melt and 55%

Table 2. Experimental run conditions and phases observed.

| | Starting composition | $fO_2$ | Run time | Phases | Average temperature | Mode glass | Mode olivine | Mode pyroxene | Mode feldspar |
|---|---|---|---|---|---|---|---|---|---|
| HAPM2_01 | PM2 | IW+1.8 | 3 | Ol, Gl | 1500 | 0.65 | 0.35 | | |
| HAPM2_02 | PM2 | IW+1.8 | 3 | Ol, Gl | 1450 | 0.54 | 0.46 | | |
| HAPM2_04 | PM2 | IW+1.8 | 3 | Ol, Gl | 1400 | 0.52 | 0.48 | | |
| HAPM2_05 | PM2 | IW+1.8 | 3 | Ol, Gl | 1350 | 0.45 | 0.55 | | |
| HAPM2_09 | PM2 | IW+1.8 | 136 | Ol, Gl, Sp | 1350 | 0.48 | 0.52 | | |
| HAPM2_12 | PM2 | IW+1.8 | 1 | Ol, Gl | 1600 | 0.94 | 0.06 | | |
| HAPM2_13 | PM2 | Air | 0.16 | Gl | 1650 | 1.00 | | | |
| HAPM2_14 | PM2 | Air | 0.16 | Gl | 1650 | 1.00 | | | |
| HAPM2_15 | PM2 | IW+1.8 | 96 | Ol, Gl, Pyx, Sp | 1300 | 0.37 | 0.53 | 0.10 | |
| HAPM2_16 | PM2 | IW+1.8 | 96 | Ol, Gl, Pyx, Sp | 1300 | 0.37 | 0.60 | 0.03 | |
| HA3_01 | 15% pyx 15% Ol 70% glass HAPM2_06 | IW+1.8 | 24 | Ol, Pyx, Gl | 1350 | 0.74 | 0.07 | 0.19 | |
| HA3_02 | 15% pyx 15% Ol 70% glass HAPM2_06 | IW+1.8 | 24 | Ol, Pyx, Gl | 1350 | 0.74 | 0.08 | 0.18 | |
| HA3_03 | 15% pyx 15% Ol 70% glass HAPM2_06 | IW+1.8 | 48 | Ol, Pyx, Gl | 1350 | 0.75 | 0.06 | 0.19 | |
| HA3_04 | 15% pyx 15% Ol 70% glass HAPM2_06 | IW+1.8 | 5 | Gl | 1575 | 1.00 | | | |
| HA3_06 | 15% Pyx 15% Ol 70% glass HAPM2_06 | IW+1.8 | 24 | Gl, Pyx, Ol | 1350 | 0.74 | 0.07 | 0.19 | |
| HA4_02 | Glass composition from HA3_06 | IW+1.8 | 24 | Glass, Pyx | 1300 | 0.84 | | 0.16 | |
| HA4_03 | Glass composition from HA3_06 | IW+1.8 | 24 | Glass | 1350 | 1.00 | | | |
| HA4_05 | Glass composition from HA3_06 | IW+1.8 | 70 | Glass, Pyx | 1250 | 0.73 | | 0.27 | |
| HA5_01 | Glass composition from HA4_05 | IW+1.8 | 66.5 | Pyx, Gl | 1150 | 0.79 | | 0.21 | |
| HA5_02 | Glass composition from HA4_05 | IW+1.8 | 70 | Pyx, Gl | 1150 | 0.80 | | 0.20 | |
| HA7_01 | Glass composition from HA5 | IW+1.8 | 60 | Pyx, Gl | 1150 | 0.96 | | 0.04 | |
| HA7_02 | Glass composition from HA5 | IW+1.8 | 60 | Pyx, Gl, Fspar | 1125 | 0.90 | | 0.09 | 0.01 |

Where the run time is in hours, the average temperature is in °C. In the phase column, olivine = ol, gl = glass, pyx = pyroxene, fspar = feldspar, sp = spinel.

olivine) and HAPM2_09 (48% melt and 52% olivine). The minor differences in phase composition are due to the higher concentrations of trace elements in HAPM2_09.

**MELTS Models**

Calculations of the crystallization path of the primitive mantle composition were performed using the alphaMELTS front-end interface of the MELTS software package (Ghiorso and Sack 1995; Smith and Asimow 2005) for comparison with the experimental data. An equilibrium crystallization model was run on the primitive mantle composition from 1700 °C down temperature at 1 °C temperature decrements and fractional crystallization models were run on the HA4 composition from 1350 °C down temperature. All models were run at atmospheric pressure and the oxygen fugacity was fixed at IW + 1.8. The MELTS model results are plotted on Figs. 1–3 for comparison with the observed experimental compositions.

**Equilibrium Partial Melting of the Primitive Mantle**

Partial melting of the mantle was simulated by taking the primitive mantle starting composition and crystallizing under equilibrium conditions.

**Crystallization Sequence**

Olivine is the liquidus phase in our putative mantle composition and was first observed at 1600 °C.

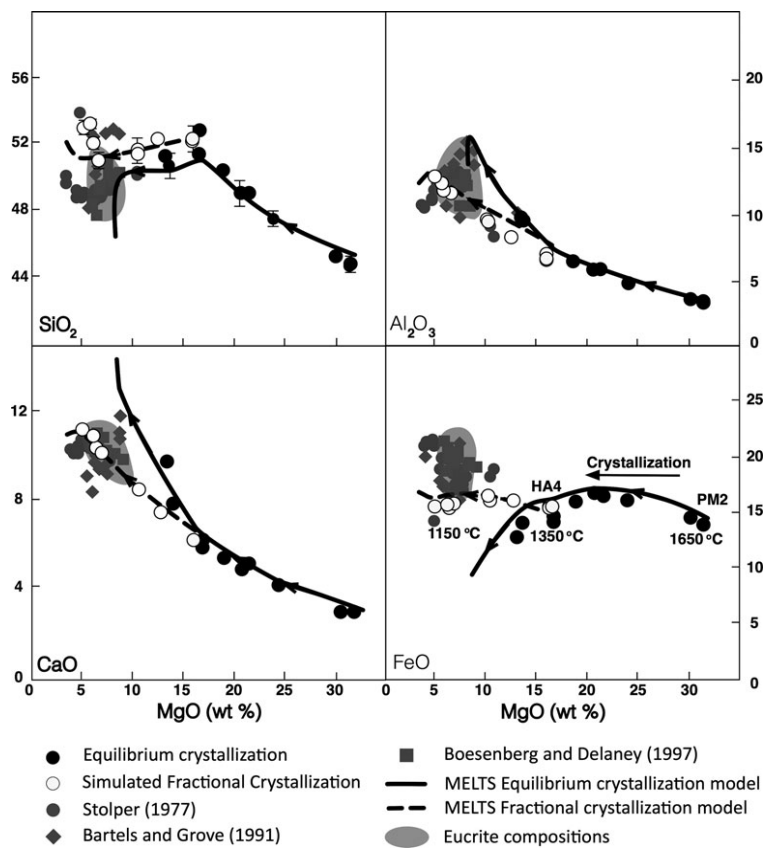

Fig. 1. Compositions of glasses from equilibrium crystallization (filled circles) and fractional crystallization (open circles) series of experiments, together with the corresponding MELTS models (lines). All results in weight%. The eucrite field covers the majority of noncumulate monomict eucrite compositions taken from Warren et al. (2009), and Barrat et al. (2000). The experimental data of Stolper (1977), Bartels and Grove (1991), and the experimental data where no significant iron loss occurred of Boesenberg and Delaney (1997) have been plotted for comparison (filled gray symbols). The arrows on the continuous lines denote the evolution of the crystallizing melt; where error bars are absent, they are smaller than the symbols.

Orthopyroxene is the next phase to crystallize, first observed at 1350 °C. Spinel is present in experiments containing $Cr_2O_3$ from 1350 °C down temperature. Only one accurate EPMA analysis of this phase was possible, however (in experiment HAPM2_15), due to its small size and low abundance. Melt compositions have been plotted on major element oxide diagrams in Fig. 1 and have also been projected into the Fo-An-Qz ternary in Fig. 2.

**Melt Compositions**

The experimental glass compositions are shown in Fig. 1 plotted as wt% $SiO_2$, $Al_2O_3$, CaO, and FeO as a function of MgO wt%. As the primitive mantle crystallizes olivine and the temperature declines from 1600 to 1300 °C the melt becomes richer in $SiO_2$ (44.68–50.63 wt%), $Al_2O_3$ (3.59–9.55 wt%), and CaO (3.06–7.87 wt%) and poorer in MgO (31.02–13.34 wt%). There is also a slight increase in $TiO_2$ over this temperature interval and FeO and $Cr_2O_3$ increase slightly as olivine crystallizes but then decrease as pyroxene and spinel join the crystalline assemblage. The Mg# of the melt decreases from 80 to 63. The melt compositions have also been projected into the Fo-An-Qz ternary in Fig. 2, with the MELTS models, and appropriate experimental melts from Stolper (1977), Bartels and Grove (1991), and Boesenberg and Delaney (1997) plotted for comparison.

**Crystal Compositions**

The Fo content (100*[Mg/(Mg+Fe)]) of olivine decreases with decreasing temperature from 92 at 1600 °C to 83 at 1300 °C. This range corresponds well to the range of olivine clasts observed in howardite GRO 95 (Lunning et al. 2015) implying that melting and crystallization of Vesta left a heterogeneous residuum. The experimental pyroxene compositions are plotted in Fig. 3. Orthopyroxene was first observed as a few very small crystals at 1350 °C. It has the composition $Wo_{1.18}En_{84.39}Fs_{14.43}$ at 1300 °C.

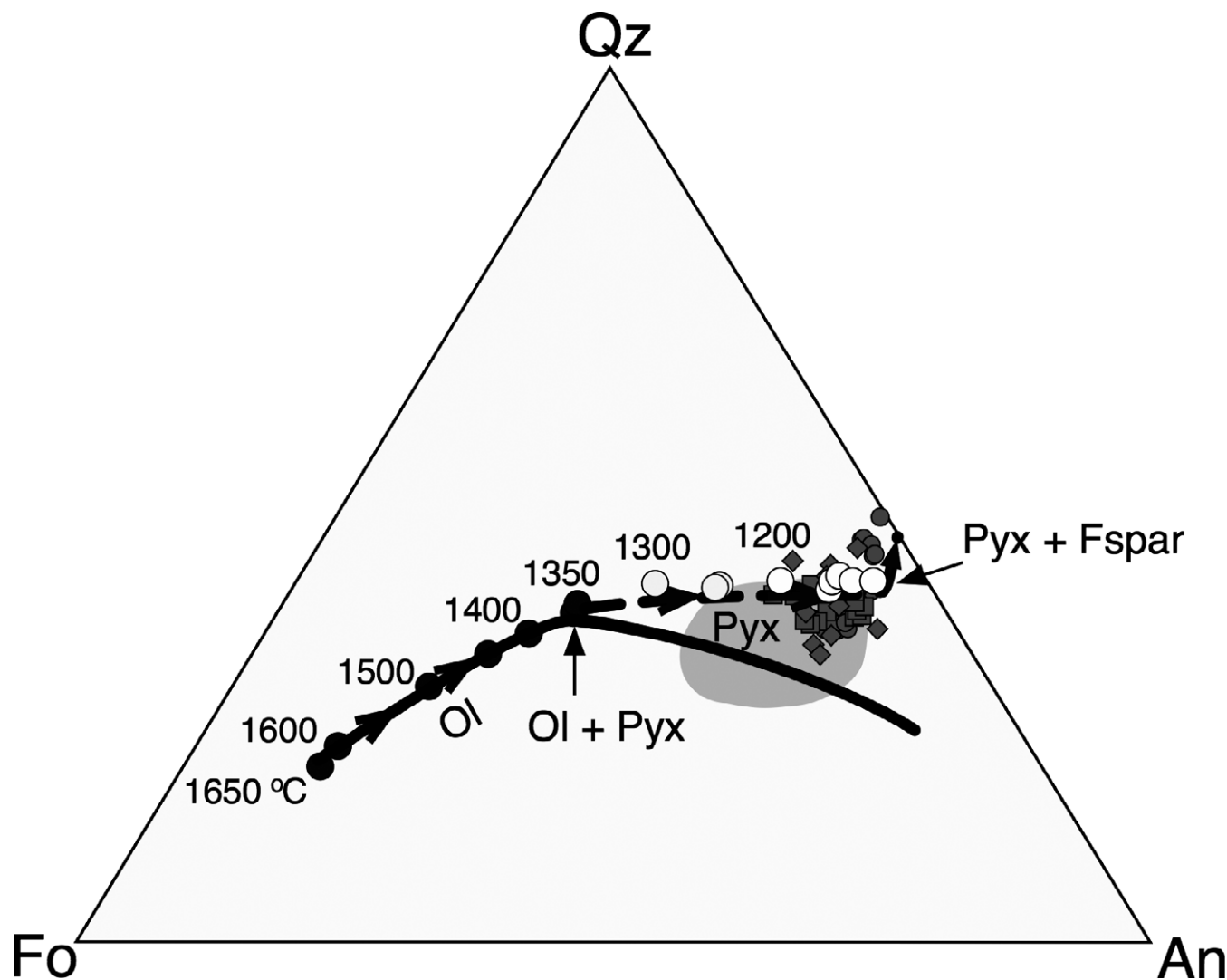

Fig. 2. A comparison of the observed and predicted melt compositions projected into the Fo-An-Qz pseudoternary from Di. Molar units are used. Legend is the same as Figure 1.

**Kd Checks for Equilibrium on Primitive Mantle**

Roeder and Emslie (1970) suggested that the olivine-liquid exchange partition coefficient $Kd_{Fe^{2+}-Mg}$ = $(FeO/MgO)_{Ol}/(FeO/MgO)_{liq}$ in terrestrial basalts is constant at a value of 0.30 ± 0.03. Later work showed, however, that $Kd_{Fe^{2+}-Mg}$ depends to some extent on liquid composition (e.g., Sack et al. 1980), and that different $Kd_{Fe^{2+}-Mg}$ values need, therefore to be used for different petrogenetic environments. In order to calculate $Kd_{Fe^{2+}-Mg}$ we need, in principle, to make a small correction for the $Fe^{3+}$ content of the liquid. We used equation 12 from Jayasuriya et al. (2004) to calculate the amount of ferric iron in the liquids generated from our primitive mantle composition at an oxygen fugacity of IW+ 1.8. After subtracting $Fe^{3+}$, a mean value of $Kd_{Fe^{2+}-Mg}$ of 0.34 ± 0.01 was obtained for our experiments. This is comparable with the values of 0.345 ± 0.009 for high MgO Hawaiian picrites (Matzen et al. 2011) and 0.35 ± 0.01 for the high FeO, low $Al_2O_3$ Martian basalts (Filiberto and Dasgupta 2011). The attainment of a constant $Kd_{Fe^{2+}-Mg}$ over the composition and temperature range studied provides further evidence for attainment of equilibrium.

**MELTS Modeling of Equilibrium Crystallization of the Primitive Mantle**

MELTS correctly predicts the observed crystallization sequence for the PM2 composition with olivine crystallizing first at 1634 °C, and orthopyroxene at 1417 °C. Experimentally, however, we find that, although the predicted olivine crystallization temperature is in good agreement with our results, the temperature of orthopyroxene appearance from MELTS is overestimated by more than 50 °C. Another known issue with the MELTS software is the way it deals with minor components of the crystalline phases. For example, no $Cr_2O_3$ is included in any other solid phase than spinel in the model. This should have some effect on the predicted liquidus of orthopyroxene because this phase generally contains some $Cr_2O_3$. It should be noted, however, that although the orthopyroxene model in MELTS is oversimplified, the lack of consideration of a Cr-bearing component is unlikely to be responsible for the significant error in its calculated liquidus temperature. This is because addition of Cr- or $Fe^{3+}$-bearing components increases the entropy of the phase and hence tends to increase rather than decrease its stability.

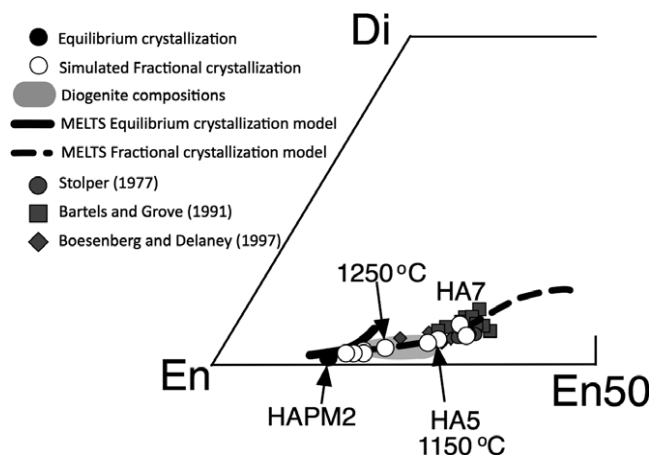

Fig. 3. Compositions of the experimental pyroxenes (filled and open circles) and the calculated compositions (lines) projected into the Wo-En-Fs ternary. The gray field corresponds to diogenite compositions taken from Fowler et al. (1995), Barrat et al. (2000), Warren et al. (2009), Mittlefehldt et al. (2012), Beck et al. (2013). The experimental pyroxenes of Stolper (1977), Bartels and Grove (1991), and Boesenberg and Delaney (1997) have been plotted for comparison.

Figures 1 and 2 show that MELTS predicts the correct trend for the liquid line of descent and melt evolution. The calculated compositions become increasingly offset from the observed compositions; however, due to the overestimation of orthopyroxene crystallization temperature, which in turn causes the MELTS program to overestimate the modal proportion of orthopyroxene and underestimate the amount of melt at any given temperature.

Despite the discrepancy noted above, the MELTS model reproduces the melt and olivine major element compositions within 1 wt% of the measured experimental compositions in the olivine phase field. Furthermore, MELTS reproduces the crystalline phase compositions at any given melt MgO content very well (on the order of 1 wt%), as illustrated in Fig. 3.

## FRACTIONAL CRYSTALLIZATION OF THE MELT EXTRACTED FROM THE MANTLE AT 1350 °C

The partial melt produced from the primitive mantle composition at 1350 °C was analyzed, synthesized, and then crystallized in a stepwise fashion down to 1125 °C in order to simulate fractional crystallization.

### Crystallization Sequence

Orthopyroxene, which appeared between 1350 and 1300 °C, was the first crystalline phase observed in the fractional crystallization sequence of liquid separated from the primitive mantle at 1350 °C. It is followed by feldspar (between 1150 and 1125 °C). Small spinel crystals are present at all temperatures in experiments with bulk compositions containing small amounts of $Cr_2O_3$.

### Melt Compositions

As fractional crystallization proceeds through the temperature interval from 1350 °C to 1125 °C the liquid composition increases in $SiO_2$ (from 51 to 53 wt%), CaO (from 8 to 11 wt%), and $Al_2O_3$ (from 9 to 13 wt%). It also decreases in MgO (from 10 to 5 wt%). The Mg# of the melt decreases from 53 at 1350 °C, to 36 at 1125 °C. The melt compositions become eucritic from 1300 °C down temperature.

### Crystalline Phase Compositions

In the fractional crystallization sequence of the melt segregated from the mantle at 1350 °C, pyroxene becomes more CaO-, FeO-, $Al_2O_3$-, and $Cr_2O_3$-rich and MgO-poor as temperature decreases, with orthopyroxene of composition $Wo_{1.31}En_{81.25}Fs_{17.44}$ at 1300 °C giving way to pigeonite below 1200 °C and the latter approaching $Wo_{6.42}En_{63.34}Fs_{30.24}$ at 1125 °C. The experimentally produced pyroxenes first become close to diogenite in composition at about 1250 °C.

### Comparison with MELTS Modeling and Meteorite Compositions

As discussed above, MELTS overestimates the temperature at which orthopyroxene crystallizes from the primitive mantle composition, but correctly predicts the temperature of feldspar crystallization in the fractional crystallization path. In the latter case MELTS predicts the correct sequence of crystallization, and reproduces the crystalline phase compositions. The errors in MELTS' predicted orthopyroxene liquidus temperature (50–100 °C) has been noted in other studies of Martian and terrestrial compositions (discussed in more detail by Balta and McSween 2013). The final experimentally determined phase proportions from the two-stage crystallization path of the putative mantle composition at 1125 °C are 24.2% melt, 52% olivine, 23.5% pyroxene, and 0.3% feldspar. MELTS predicts similar proportions of 18.8% melt, 48.9% olivine, 32.1% pyroxene, and 0.15% feldspar.

To compare the models directly with the HED meteorites, minor amounts of $Na_2O$ (0.11 wt%), and $K_2O$ (0.01 wt%) were added to the primitive mantle composition, and MELTS calculations were repeated at IW −1 (following Mandler and Elkins-Tanton 2013). The calculated crystallization sequence at IW −1 is virtually identical to that calculated at IW +1.8 which indicates that our experiments performed at IW +1.8 are appropriate representations of eucrite crystallization. MELTS calculations at slightly elevated pressures of 500 bars were also performed on the alkali-bearing primitive mantle composition. In this case, a few percent of olivine crystallizes before orthopyroxene in the fractional crystallization part of the model, but aside from this, the crystallization sequence and phase compositions are essentially the same. In terms of phase proportions, the MELTS algorithm predicts that 55% equilibrium crystallization (45% partial melting) and subsequent fractional crystallization of our primitive mantle composition would yield a mantle which is 85% olivine, 15% orthopyroxene, and a crust which is 50:50 eucrite:diogenite material, consistent with the proportions of eucritic and diogenitic meteorites.

Previous experimental studies relevant to eucrite genesis are those of Stolper (1977) who determined the phase relations of two main group eucrites, Juvinas and Sioux County, in the liquidus–solidus interval at atmospheric pressure; Bartels and Grove (1991), who performed melting experiments on two magnesium-rich eucrite clasts, Kapoeta and Yamato 7308; and Boesenberg and Delaney (1997), who performed equilibrium melting experiments on a 70% H—30% CM chondrite mixture to produce eucritic liquids. The compositions of melts generated in these studies at appropriate temperatures and pressures are plotted for comparison with ours on Figs. 1–3. The melts produced by 45% batch partial melting of our primitive mantle composition followed by fractional crystallization of the 1350 °C melt are compositionally similar to those generated in the earlier studies. They tend, however, to be slightly richer in $SiO_2$ and poorer in FeO. The most evolved pyroxene compositions we produce, however, match the low-Ca pyroxene compositions found in eucrites and in the Stolper (1977), Bartels and Grove (1991), and Boesenberg and Delaney (1997) experiments.

Our primitive mantle composition (Table 1) with chondritic ratios of refractory lithophile elements is too olivine-rich and has too high an Mg# to generate eucritic liquids and diogenitic pyroxenes directly through equilibrium crystallization (Fig. 2). However, if a 45% equilibrium partial melt of such a mantle were extracted then fractional crystallization of the melt would generate eucritic melts and diogenitic pyroxenes at temperatures of 1250 °C and below (Figs. 1–3). The primitive mantle composition derived here is compositionally similar to the 0.3 CV—0.7 L mixture of CV and L type chondrites that Righter and Drake (1997) favor, but has a higher Mg# of 80 compared to their value of 74. In the latter case, optimal eucrite and diogenite compositions should be generated after about 60–70% crystallization (Mandler and Elkins-Tanton 2013). As both compositions can produce eucrites and diogenites, there is obviously a range of acceptable mantle compositions. We therefore used our composition of Vesta's mantle (PM2) as a starting point for further investigation of the possible range of mantle compositions and crystallization sequences.

## DISCUSSION OF PRIMITIVE MANTLE COMPOSITION

The Mg# of the starting mantle and the extent of equilibrium crystallization prior to melt extraction were varied systematically between 65 and 85, and 50% to 90%, respectively, in a series of MELTS calculations in order to determine the range of mantle compositions capable of producing eucrites and diogenites. The relative proportions of the other major elements were kept chondritic. The calculation was considered successful if during the fractional crystallization path, the melt evolved into the eucrite composition field and orthopyroxenes of diogenite composition could crystallize independently of other solid phases.

As the Mg# increases from 0.65 to 0.85 the olivine liquidus temperature increases from 1617 to 1661 °C. Orthopyroxene is the second phase to crystallize in mantle compositions with Mg# ≥70 with the liquidus increasing from 1319 °C to 1460 °C. For comparison with the experimental work here, melts extracted at 1350 °C can be compared. The extent of crystallization varies from 55 to 71% as Mg# increases from 65 to 85, and the Mg# of the extracted melt increases from 44 to 72. This affects the subsequent mineralogy of the Vestan crust, with the proportion of pyroxene to olivine crystallized from the extracted melt increasing with starting mantle Mg#. The starting mantle composition with a Mg# of 85 does not crystallize olivine in addition to pyroxene, and the extracted melt from the mantle composition with Mg# 65 does not crystallize pyroxene in addition to olivine. An essential requirement for the formation of the diogenites through a two-stage model is that the extracted partial melts crystallize dominantly pyroxene and have a high pyroxene:olivine ratio during fractional crystallization. The extracted melt from the Mg# 70 mantle crystallizes equal proportions of olivine and pyroxene so a lower limit of Mg# ~75 can be placed on the starting mantle compositions. If we assume that the proportions of refractory elements, excluding Fe, in the mantle are chondritic, then melts in the range of eucrite compositions are produced from starting compositions with Mg# of 75–80 through the magma ocean model. If the Mg# is higher, the melts are more $SiO_2$ rich and FeO poor than the eucrite field, and if the starting Mg# is lower the melts are lower in $SiO_2$, CaO, and $Al_2O_3$ and more FeO-rich than the eucrite composition field. Therefore, assuming development of an early magma ocean, the plausible range of Mg# for the starting Vestan mantle is ~ 75–80.

The alternative, partial melting model can yield eucritic melts by 20% partial melting of a mantle of Mg# ~70. In such cases, however, the residuum would be olivine-rich and the accumulating pyroxene too low in Mg# to represent the most magnesian diogenites. More magnesian orthopyroxenes of Mg# 80 could be produced by crystallization from melts extracted after 30–50% partial melting but the fractionating melts do not become eucritic, and there is no appropriate lithology in the meteorite record. The only plausible way, therefore, in which both eucrites and diogenites could be produced from Mg # 70 mantle, is if this source region were heterogeneous or if it underwent successive melting and crystallization events. Such explanations would be in accord with the observed wide range of trace element abundances including the Ni and Co contents of Mg-rich clasts (Lunning et al. 2015) and Dy/Yb ratios of diogenites (Barrat and Yamaguchi 2014).

In order to investigate how the extent of partial melting prior to melt extraction affects the liquid–solid relationships, the fractional crystallization path of partial melt of our Mg# 80 primitive mantle determined experimentally at 1300 °C (after 37% partial melting) was calculated using MELTS. The extracted melt and its crystallization path generates liquids higher in $SiO_2$, CaO, and $Al_2O_3$ with lower FeO contents than the partial melt at 1350 °C (after 45% melting). The liquid line of descent is not significantly different, however, and eucritic liquids and diogenitic pyroxenes are both produced. We are aware that the error in the orthopyroxene liquidus calculated by MELTS may lead to a slight overestimate of the degree of partial melting required to generate eucrites from any given mantle composition. However, the general agreement between our experimentally produced melt compositions and those calculated by MELTS means that the effects of changing bulk composition given by MELTS should be reasonably accurate.

For starting mantle compositions with Mg# of ~ 80, melt extraction after 35–45% partial melting is required to generate eucritic liquids. The partial melts produced are in the eucrite field, and continue to evolve as the MgO content decreases and crystallization proceeds. For lower degrees of partial melting the melt is not in the eucrite field and is too poor in FeO and rich in $SiO_2$ to be able to generate eucrites. For starting mantle compositions with Mg# of 75, a slightly lower degree of partial melting (30–40%) produces a partial melt which is eucritic in composition and can continue to evolve in the eucrite field. If melt is extracted at a higher temperature/degree of partial melting the melt composition is not within the eucrite field and is too FeO-rich, CaO- and $Al_2O_3$-poor to become eucritic while low degrees of partial melting (<30%) leads to melts which are too poor in MgO and FeO to produce eucrites.

Finally, a more general investigation into the primitive mantle composition was performed using a Monte-Carlo approach. Starting compositions were generated by randomly selecting the concentrations of each major element (MgO, FeO, CaO, $Al_2O_3$, $SiO_2$) within a ±10% relative window about the primitive mantle composition. These values were then normalized to 100% to produce a new bulk mantle composition. These ±10% variations in MgO, FeO, CaO, and $Al_2O_3$ do not significantly affect the crystallization sequence and phase proportions; however, we do find that a starting mantle with >43% $SiO_2$ is required to generate orthopyroxene during fractional crystallization.

The results from the experimental and modeling work performed here indicate that the different petrogenetic models require different bulk silicate Vesta compositions given the underlying assumption that CaO, MgO, $Al_2O_3$, and $SiO_2$ are in chondritic proportions to one another. A range of HED mantle compositions between our PM2 (Table 1) with Mg# 80, and the 0.3 CV—0.7 L composition of Righter and Drake (1997) with Mg# 75 can produce the major element compositions of both eucrites and diogenites through the two-stage model of partial melting, melt extraction, and fractional crystallization. Eucritic crustal compositions (but not the full range of diogenite precipitates) can also be produced by partial melting of a HED mantle composition with a slightly lower Mg# of 70. The latter is similar in composition to the mixture of H and CM chondrites favored by Boesenberg and Delaney (1997) and Toplis et al. (2013).

This experimental study has been aimed at constraining the general relationships between the bulk composition of silicate Vesta and those of eucrites and diogenites. To this end we have presented a simplified petrogenetic scheme based on the magma ocean model. We do not assert that this provides a complete magmatic history of Vesta, however. In detail, Vesta is more complex, as indicated by trace element abundances in eucrites and diogenites. If, for example, all eucrites and diogenites were related to one another by fractional crystallization, the Dy/Yb ratio of diogenites would always be in simple ratio to that in eucrites. This ratio would be given by the ratio of partition coefficients into low-Ca pyroxene, which is ~0.5 (Barrat and Yamaguchi 2014). The observed range of chondrite-normalized Dy/Yb in diogenites is 0.25–0.89, however, indicating a wide range of REE concentrations in their sources. In contrast, eucrites exhibit relatively flat chondrite-normalized REE patterns with a small range in Dy/Yb. The implication is that, while fractional crystallization of the eucrites produces diogenitic pyroxenes in terms of major element compositions, the magmatic history of diogenites is more complex than required by a single episode of fractional crystallization. In order to generate the observed range of REE contents, the diogenitic cumulates must have undergone remelting and possibly partial re-equilibration subsequent to eucrite fractionation (Barrat and Yamaguchi 2014). This added complexity is not inconsistent with the general petrogenetic scheme we have presented, however, and is, we believe, simply reflective of the complex history to be expected on a large body.

## CONCLUSIONS

We find that a HED parent body with refractory elements (including Fe, Mg, and Si) in chondritic proportions results in a model consistent with the density and moment of inertia of Vesta if the mantle has an Mg# of about 80. In this case, the core would be 15–20% of the mass of Vesta and the mantle would contain olivine of approximately the same Mg# as the most Mg-rich diogenites. The liquidus temperature of the mantle is ~1625 °C with olivine being the sole precipitating phase in the interval 1625–1350 °C. Separation of the mantle melt at 1350 °C, the point at which orthopyroxene appears, simulates equilibrium partial melting of the mantle and ensures that the liquid line of descent is dominated by pyroxene fractionation. Under these conditions, 45% equilibrium partial melting of the mantle leads to a melt which, during fractional crystallization, produces eucritic liquids and diogenitic solid assemblages at temperatures below ~1300 °C. We thus find that our mantle composition is in the acceptable range for the generation of the HED meteorites by equilibrium partial melting followed by fractional crystallization of the segregated melt (Righter and Drake 1997).

We found that the MELTS program performs well at predicting the olivine liquidus temperature, overestimates the orthopyroxene liquidus by ~70 °C, and predicts melt compositions in good agreement with those observed at any given MgO content. Given its utility, we used the MELTS program to investigate the range of mantle compositions which can lead to eucritic melt compositions and which precipitate pyroxenes of diogenitic composition. We find that, given the assumption of chondritic ratios of refractory lithophile elements, the range of mantle Mg#s which generate eucrites and diogenites by the two-stage process is 75–80. A Monte Carlo approach suggests that small variations (±10% relative to their major element abundance) in CaO, MgO, and $Al_2O_3$ content of the HED parent body mantle remain consistent with generation of eucrites (liquids) and diogenites but that the $SiO_2$ content of the mantle must be greater than 43 wt% in order to generate orthopyroxene during fractional crystallization of the segregated partial melts. Although the magma ocean model generates a good approximation to the compositions of eucrites and diogenites by equilibrium partial melting followed by fractional crystallization, it is apparent that the range of solid compositions requires more complexity of process. Incomplete re-equilibration, re-melting of cumulates, and differential melt segregation during cooling are probably required to match the observed compositional ranges in terms of incompatible trace elements.

*Acknowledgments*—The authors thank A. Yamaguchi, J. A. Barrat, and K. Righter for their constructive reviews and suggestions. H. A. acknowledges the support of a studentship from the Science and Technology Facilities Council (UK).

*Editorial Handling*—Dr. Akira Yamaguchi

## SUPPORTING INFORMATION

Additional supporting information may be found in the online version of this article:

**Table S1**: Major element compositions of experimental products.